\documentclass[preprint]{aastex631}

\usepackage{enumitem,amssymb}
 \usepackage{amsmath}
\newlist{todolist}{itemize}{2}
\setlist[todolist]{label=$\square$}
\usepackage{pifont}

\usepackage{soul}
\usepackage{colortbl}
\usepackage{multirow}
\graphicspath{{./}{figures/}}

\received{October 18, 2021}
\revised{November 17, 2021}
\accepted{December 1, 2021}
\submitjournal{PSJ}

\shorttitle{Modeling Meteoroid Impacts on the Juno spacecraft}
\shortauthors{Pokorn\'{y} and et al.}

\begin{document}

\title{Modeling Meteoroid Impacts on the Juno spacecraft}

\correspondingauthor{Petr Pokorn\'{y}}
\email{petr.pokorny@nasa.gov}

\author[0000-0002-5667-9337]{Petr Pokorn\'{y}}
\affiliation{Department of Physics, The Catholic University of America, Washington, DC 20064, USA}
\affiliation{Astrophysics Science Division, NASA Goddard Space Flight Center, Greenbelt, MD 20771}
\affiliation{Center for Research and Exploration in Space Science and Technology, NASA/GSFC, Greenbelt, MD 20771}

\author[0000-0003-2685-9801]{Jamey R. Szalay}
\affil{Department of Astrophysical Sciences, Princeton University, 4 Ivy Ln., Princeton, NJ 08540, USA}

\author[0000-0002-5920-9226]{Mih\'aly Hor\'anyi}
\affiliation{Department of Physics, University of Colorado Boulder, 392 UCB, Boulder, CO, 80309, USA}
\affiliation{Laboratory for Atmospheric and Space Physics, 1234 Innovation Dr., Boulder, CO, 80303, USA}
\affiliation{Institute for Modeling Plasma, Atmospheres, and Cosmic Dust, 3400 Marine St., Boulder, CO, 80303, USA}

\author[0000-0002-2387-5489]{Marc J. Kuchner}
\affiliation{Astrophysics Science Division, NASA Goddard Space Flight Center, Greenbelt, MD 20771}

\begin{abstract}
Events which meet certain criteria from star tracker images onboard the Juno spacecraft have been proposed to be due to interplanetary dust particle impacts on its solar arrays. These events have been suggested to be caused by particles with diameters larger than 10 micrometers. Here, we compare the reported event rates to expected dust impact rates using dynamical meteoroid models for the four most abundant meteoroid/dust populations in the inner solar system. 
{We find that the dust impact rates predicted by dynamical meteoroid models are not compatible with either the Juno observations  in terms of the number of star tracker events per day, or with the variations of dust flux on Juno's solar panels with time and position in the solar system.}
For example, the rate of star tracker events on Juno's anti-sunward surfaces is the largest during a period during which Juno is expected to experience the peak impact fluxes on the opposite, sunward hemisphere. We also investigate the hypothesis of dust leaving the Martian Hill sphere originating either from the surface of Mars itself or from one of its moons. We do not find such a hypothetical source to be able to reproduce the star tracker event rate variations observed by Juno. We conclude that the star tracker events observed by Juno are unlikely to be the result of instantaneous impacts from the Zodiacal Cloud. 
\end{abstract}

\section{Introduction}
In-situ detections of dust and meteoroid impacts larger than several micrometers are extremely rare. This is due to the tenuous nature of the Zodiacal Cloud where we can expect approximately 1 impact a day for particles larger than $D=2.5~\mu$m on a randomly spinning flat 1 m$^2$ detector orbiting the sun at 1 au \citep{Grun_etal_1985}. Only a handful of space missions were able to accumulate enough time and collecting area to provide a sizeable dataset such as those from Pegasus \citep{Clifton_etal_1966} and the Long Duration Exposure Facility \citep[LDEF;][]{Love_Brownlee_1993}. However, these missions were on low Earth orbits and did not probe the meteoroid flux outside Earth's Hill sphere. Moreover, the correct interpretation of these data sets is still debated and continuously re-analyzed \citep{Moorhead_etal_2020}.

It has been proposed that Juno's 60 m$^2$ solar arrays, in combination with star tracker cameras, can be used as a dust impact detector \citep{Benn_etal_2017}. When a dust particle/meteoroid impacts the solar panel with $>1$ km s$^{-1}$ impact velocity, it creates an ejecta cloud composed of a mixture of the impactor and the solar panel surface material. If the ejected debris is bright and large enough, it could be detected with star tracker cameras, and due to its unique trajectory, separated from background objects such as stars and asteroids \citep{Benn_etal_2017}. The detection method is restricted to the anti-sunward facing side of the solar arrays because the star tracker cameras would be blinded by the sunward facing part of the solar array. The first analysis reported 13 star tracker events (STEs) that were proposed to be due to distinct dust impacts during 3.5 years of Juno's travel to Jupiter from January 2013 to June 2016  \citep{Benn_etal_2017}. 

An extended analysis {\added{of \citet{Jorgensen_etal_2021}} found \added{a total of }} 15,278 STEs. {\added{These STEs were}} proposed to be due to interplanetary dust particle impacts the solar arrays{\added{, where the increase in the number of STEs was}} due to more in-depth analysis of the entire Juno star tracker data set \citep{Jorgensen_etal_2021}. A series of filters were applied  to isolate interplanetary dust particle (IDP) impacts from other luminous sources such as stars, planets, and asteroids resulting in a lower limit of detected IDPs \citep{Jorgensen_etal_2021}. The impact rate profile inferred from the STEs between Earth and Jupiter was unexpected and remain{\replaced{s}{ed}} at odds with established models of both the asteroid dust bands \citep{Nesvorny_etal_2006,Nesvorny_etal_2010} and the inner Zodiacal Cloud \citep{Nesvorny_etal_2010,Nesvorny_etal_2011JFC,Nesvorny_etal_2011OCC,Pokorny_etal_2014}{\added{; Juno's STEs were according to \citet{Jorgensen_etal_2021} dynamically linked to IDPs originating from 5:1 (1.779 au) and 4:1 (2.064 au) mean-motion resonances with Jupiter, whereas established dust models place their source populations at and beyond the main belt} }. Interpreting the STEs to be due to individual dust impacts, a new simplified model for the dust and meteoroid environment from two distinct dust populations was proposed : (1) a primary dust population associated with Mars and sharing its orbital elements, and (2) a secondary population that results from scattering the primary population through Kozai-Lidov oscillations \citep{Jorgensen_etal_2021}. However, neither the source mechanism for these two dust populations, nor an explanation for the disappearance of all currently known dust sources (asteroids and comets) have been presented in the literature to date.

\section{Methods}
In this Section we first review  the traditional meteoroid models used in this article to supply the position and velocity vector distributions of dust grains in the inner solar system. We also  introduce a simple dynamical model for a putative dust cloud generated from dust escaping Mars' Hill sphere.
We discuss the collision probability between particles from the dust cloud and Juno during its flight. Ultimately, we analyze Juno's trajectory and pointing of its solar arrays that are crucial for the correct interpretation of our modeling efforts. 

\subsection{Meteoroid models for the Zodiacal Cloud}
In this article, we use a four population dynamical meteoroid model  \citep{Pokorny_etal_2019,Pokorny_etal_2020}. This model combines dust and meteoroids generated from the four most abundant sources of dust in the inner solar system: main belt asteroids \citep{Nesvorny_etal_2010}, Jupiter-family comets \citep{Nesvorny_etal_2011JFC}, Halley-type comets \citep{Pokorny_etal_2014}, and Oort Cloud comets \citep{Nesvorny_etal_2011OCC}. The particle diameter range for all models is $D=10-2000~\mu$m, where for all populations we assume the same bulk density $\rho=2000$ kg m$^{-3}$. There are several free parameters for the dynamical meteoroid model used here: the size-frequency distribution at the source following a single power-law with a differential size index $\alpha$, the collision probability multiplier $F_\mathrm{coll}$, and the mass accreted at Earth $M_\mathrm{pop}$ for each of the four populations. The basic summary of all four meteoroid population models used here and their free parameters are shown in Table \ref{TABLE:Model_Description}. {\added{Models used in this article were constrained by numerous inner solar system observations such as Infrared Astronomical Satellite (IRAS) observations of the Zodiacal Cloud \citep{Low_etal_1984,Hauser_etal_1984, Nesvorny_etal_2010}, orbital distributions of radar meteors at Earth \citep{CampbellBrown_2008,Galligan_Baggaley_2004}, meteoroid mass flux at Earth \citep{Love_Brownlee_1993,CarrilloSanchez_etal_2016,CarrilloSanchez_etal_2020}or meteor size-frequency distribution at Earth \citep{Grun_etal_1985} and used to explain or reproduce various meteoroid related phenomena on Mercury \citep{Pokorny_etal_2017_Mercury, Pokorny_etal_2018}, Venus \citep{Janches_etal_2020}, Earth \citep{Swarnalingam_etal_2019}, Moon \citep{Janches_etal_2018, Pokorny_etal_2019}, Mars \citep{CarrilloSanchez_etal_2020}, or Ceres \citep{Pokorny_etal_2021}
}}
\begin{deluxetable}{llcl|lc}

\tablecaption{\label{TABLE:Model_Description}Description of meteoroid dynamical models used in this work. The meteoroid model used here has six free parameters: the collisional lifetime multiplier $F_\mathrm{coll}$ \citep{Pokorny_etal_2014}, differential size-frequency index $\alpha$, and the average daily mass influx at Earth $M_\mathrm{pop}$ for each of the four populations in metric tons per day (1000 kg per day or 11.57 g s$^{-1}$) \citep{CarrilloSanchez_etal_2016,Pokorny_etal_2019}.  \added{The total number of particle records $N_\mathrm{rec}$ in the meteoroid models used in this article are: MBA: $N_\mathrm{rec} = 462\times 10^6$, JFC: $N_\mathrm{rec} = 305\times 10^6$, HTC: $N_\mathrm{rec} = 327\times 10^6$, OCC: $N_\mathrm{rec} = 259\times 10^6$}. For more detailed information refer to references in the table or \citet{Pokorny_etal_2018}.}

\tablenum{1}

\tablehead{\colhead{Source population} & \colhead{Acronym} & \colhead{Diameter} & \colhead{Reference} & \colhead{Mass influx at} & \colhead{Parameter}\\
& &  {($\mu$m)} & &Earth (tons/day)& Settings
} 


\startdata
Main-belt asteroids & MBA & 10 -- 2000 & \citet{Nesvorny_etal_2010} & $M_\mathrm{MBA}=3.7$ & $F_\mathrm{coll}=20$ \\
Jupiter-family comets & JFC & 10 -- 2000 & \citet{Nesvorny_etal_2011JFC} & $M_\mathrm{JFC}=34.6$ &  $\alpha=-4.0$ \\
Halley-type comets & HTC & 10 -- 2000 & \citet{Pokorny_etal_2014}& $M_\mathrm{HTC}=2.82$ & \\
Oort Cloud comets & OCC & 10 -- 2000 & \citet{Nesvorny_etal_2011OCC} & $M_\mathrm{OCC}=2.12$  &\\ 
\enddata




\end{deluxetable}

\subsection{Collision probability between the spacecraft and the dust particle cloud}
\label{SEC:Collisional_Probability}
The meteoroid cloud in our dynamical models is represented as a list of particle records. For each model particle we know the diameter $D$, number of meteoroids it represents $N_\mathrm{met}$, and the six orbital elements $(a,e,i,\Omega,\omega,M)$, where $a$ is the semimajor axis, $e$ is the eccentricity, $i$ is the orbital inclination, $\Omega$ is the longitude of the ascending node, $\omega$ is the argument of pericenter, and $M$ is the mean anomaly. For the spacecraft itself, we use the SPICE framework to obtain the mean daily position and velocity vector of Juno $(\vec{r}_\mathrm{Juno}, \vec{v_\mathrm{Juno}})$, and the pointing of the solar array in heliocentric ecliptic coordinates $({\lambda_\mathrm{Juno}-\lambda_\odot}, \beta_\mathrm{Juno})$,{ \added{where $\lambda_\odot$ is the ecliptic longitude of the sun, solar longitude}}. This provides sufficient information to estimate the number of modeled meteoroid impacts on Juno's solar arrays.

 We estimate the impact probability $\mathcal{P}$ of each particle record in the model dust clouds with the Juno spacecraft using the particle orbital elements $(a,e,i)$ and spacecraft {\added{state vector}} $(\vec{r}_\mathrm{Juno}, \vec{v_\mathrm{Juno}})$. The probability of a collision between a particle and the spacecraft {\added{per unit time}} is
\begin{equation}
    \mathcal{P} = \frac{V_\mathrm{rel} \sigma}{2\pi^3 R_\mathrm{hel} a \left[ \left ( \sin^2i - \sin^2\beta\right)\right]^{1/2} \left[ \left(R_\mathrm{hel} - q \right) \left( Q - R_\mathrm{hel} \right) \right]^{1/2} },
\end{equation}
where $V_\mathrm{rel}$ is the relative impact velocity between the particle and the spacecraft, $\sigma$ is the collision cross-section, $R_\mathrm{hel} = ||\vec{r_\mathrm{Juno}} || $ is the heliocentric distance of the collision/spacecraft, $\beta = \mathrm{asin}(z_\mathrm{Juno}/R_\mathrm{hel})$ is the ecliptic latitude of the spacecraft at the time of the collision, and $q$ and $Q$ are the meteoroid pericenter and apocenter distance, respectively \citep{Kessler_1981}.

The calculation of the relative impact velocity $V_\mathrm{rel}$ from $(a,e,i)$ is also readily available \citep{Kessler_1981}. Alternatively, we can use $\vec{r}_\mathrm{Juno}$ and $(a,e,i)$ to derive the orbital velocity vector of the particle record $\vec{v}_\mathrm{par}$ at $\vec{r}_\mathrm{Juno}$ and then $V_\mathrm{rel} = || \vec{v}_\mathrm{rel}|| = || \vec{v}_\mathrm{Juno} - \vec{v}_\mathrm{par} ||$. Having $\vec{v}_\mathrm{rel}$ allows us to calculate the heliocentric ecliptic coordinates of impacting particles  $({\lambda_\mathrm{par}-\lambda_\odot}, \beta_\mathrm{par})$. Using the known orientation of the spacecraft $({\lambda_\mathrm{Juno}-\lambda_\odot}, \beta_\mathrm{Juno})$ and great-circle distance {\added{formula}}, we can determine at what incident angle $\varphi$ the meteoroids impact the solar arrays
{\begin{equation}
    \cos\varphi = \sin\beta_\mathrm{Juno} \sin{\beta_\mathrm{par}} +  \cos\beta_\mathrm{Juno} \cos\beta_\mathrm{par} \cos{| \lambda_\mathrm{Juno} - \lambda_\mathrm{par}| }. 
\end{equation}
\added{Then, w}e} calculate the collision cross-section $\sigma$ as

\begin{equation}
    \sigma = A_\mathrm{Juno} \cos{\varphi},
    \label{EQ:Collecting_Area}
\end{equation}
where $A_\mathrm{Juno} = 60$ m$^{2}$ is the collecting area of Juno's solar arrays.
 The detection efficiency of the initial analysis was estimated to be   $\epsilon = 0.07$ for impactors tens of micrometers in diameter \citep{Benn_etal_2017}. For the purpose of this article, we assume that all meteoroid impacts with impactor $D>10~\mu$m are detected with 7\% efficiency and smaller impactors do not produce detectable STEs. Compared to the earlier studies  \citep{Benn_etal_2017, Jorgensen_etal_2021}, our $D>10~\mu$m threshold provides an upper limit on the modeled number of meteoroid impacts. This allows us to revisit the detection efficiency or the model parameters should the number of modeled impacts exceed the values reported for Juno's solar array \citep{Jorgensen_etal_2021}.

Ultimately, the number of STEs expected to be detected by Juno per day, assuming heliocentric position and velocity vectors $(\vec{r}_\mathrm{Juno}, \vec{v}_\mathrm{Juno})$ and pointing $({\lambda_\mathrm{Juno}-\lambda_\odot}, \beta_\mathrm{Juno})$, is
\begin{equation}
    N(\vec{r}_\mathrm{Juno}, \vec{v}_\mathrm{Juno},{\lambda_\mathrm{Juno}-\lambda_\odot}, \beta_\mathrm{Juno}) = \sum_k \mathcal{P}(a_k,e_k,i_k,\vec{r}_\mathrm{Juno}, \vec{v}_\mathrm{Juno},{\lambda_\mathrm{Juno}-\lambda_\odot}, \beta_\mathrm{Juno}) N_\mathrm{met_k} \epsilon t,
    \label{EQ:IMPACT_RATE}
\end{equation}
where $k$ is the index of a meteoroid model particle record, and $t=86400$ seconds is the time period over which the collisions are happening. We sum over all records in the meteoroid model.

\subsection{Hypothetical dust cloud generated from the Martian system}
\label{SEC:MARS_MODEL}
A large portion of the STEs observed by Juno have been suggested to be caused by dust  produced near the orbit of Mars or by Mars itself \citep{Jorgensen_etal_2021}. We simulate the dust generated inside Mars' Hill sphere by either Mars or its satellites Phobos and Deimos, through generating dust particles at the edge of the Martian Hill sphere by assigning the particles' position vector 

\begin{equation}
\vec{r_\mathrm{par}} = \vec{r_\mathrm{Mars}} + ||\vec{n_\mathrm{rand}}|| R_\mathrm{hill},
\label{EQ:MARS_POS}
\end{equation}
where $\vec{r_\mathrm{Mars}}$ is the position vector of Mars at the time of particle ejection, $||\vec{n_\mathrm{rand}}||$ is the normalized vector generated as $(\cos{\phi}\sin{\theta}, \sin{\phi}\sin{\theta}, \cos{\theta}$) with $\phi$ randomly selected from $[0,2\pi)$ and $\theta$ randomly selected from $[-\pi/2,\pi/2]$, and $R_\mathrm{hill} = 0.0066$ au is the radius of the Martian Hill sphere. All particles generated upon ejection are given a velocity kick $V_\mathrm{kick}$ pointing randomly so the velocity vector of the ejected particle is
\begin{equation}
\vec{v_\mathrm{par}} = \vec{v_\mathrm{Mars}} + ||\vec{n_\mathrm{rand}}|| V_\mathrm{kick},
\label{EQ:MARS_VEL}
\end{equation}
where $\vec{v_\mathrm{Mars}}$ is the velocity vector of Mars at the time of particle ejection, and $0 \le V_\mathrm{kick} \le 1$ km s$^{-1}$. {\added{We only consider particles with velocity vectors pointing outside Mars' Hill sphere; i.e. we replace any inward pointing particles with their outward pointing randomly generated counterparts.}}

For all generated particles, we assume the bulk density $\rho_\mathrm{Mars} = 1500$ kg m$^{-3}$, i.e. a value similar to that observed at the surface of Mars \citep{Moore_etal_1999}. In our dust cloud simulation, we track the dynamical evolution of particles having 7 different diameters $D=1.5, 2.0, 2.5, 5.0, 10.0, 25.0 ,50.0~\mu$m. For each size we generate 5,000 particles using Eqs. \ref{EQ:MARS_POS} and \ref{EQ:MARS_VEL}. Due to solar radiation, micron-sized particles are blown out of the solar system on hyperbolic orbits. The critical heliocentric distance for ejection on hyperbolic orbits is
\begin{equation}
R_\star = 2\beta a,
\end{equation}
where $\beta = 1.15\times10^{-3}/(\rho_\mathrm{par}D)$ is the ratio between the radiative and gravitational force \citep{Burns_etal_1979} and $\rho_\mathrm{par}$ and $D$ are in MKS units, i.e. for our $D=1.5~\mu$m particle we get $\beta = 0.511$. Particles smaller than $D=1.5~\mu$m are ejected on hyperbolic orbits regardless of the time of ejection from Martian Hill's sphere due to the low eccentricity of Mars; $R_\star < a_\mathrm{Mars}(1-e_\mathrm{Mars})$. Particles with $D=2.0, 2.5~\mu$m are ejected on both bound and unbound orbits depending on the time of their ejection. Particles with $D\ge5.0~\mu$m are always ejected on bound orbits.

Particles larger than $D=50.0~\mu$m are expected to have similar dynamical pathways as particles with $D=50~\mu$m. We assume, based on the number density of Zodiacal Cloud meteoroids \citep{Grun_etal_1985}, that these larger particles ($D>50~\mu$m) have orders of magnitude smaller spatial number density than their smaller counterparts and will not significantly contribute to the number of observed impacts on Juno's solar arrays.

Our model dust particles are ejected at 10 different positions of Mars uniformly spaced in time by 68.5 days, with the first position starting on January 1st, 2000 at 12:00 UTC. We thus create 10 distinctive dust clouds and let them evolve in time until all particles impact one of the planets, are closer than 0.05 au to the sun, or are farther than 10,000 au from the sun. All particles are numerically integrated using the \texttt{SWIFT\_RMVS\_3} numerical integrator \citep{Levison_Duncan_2013}, where the effects of the Poynting-Roberson drag and radiation pressure are included \citep{Burns_etal_1979}. The effect of the solar wind on the particle dynamics is included as a 30\% enhancement of the magnitude of the Poynting-Robertson drag \citep{Mukai_Yamamoto_1982,Gustafson_1994}. We do not take into account particle collisions with the Zodiacal Cloud to maximize the potential contribution  of the hypothetical Martian dust cloud.

\subsection{Juno's trajectory and pointing}
\label{SEC:Trajectory}
\begin{figure}
\plotone{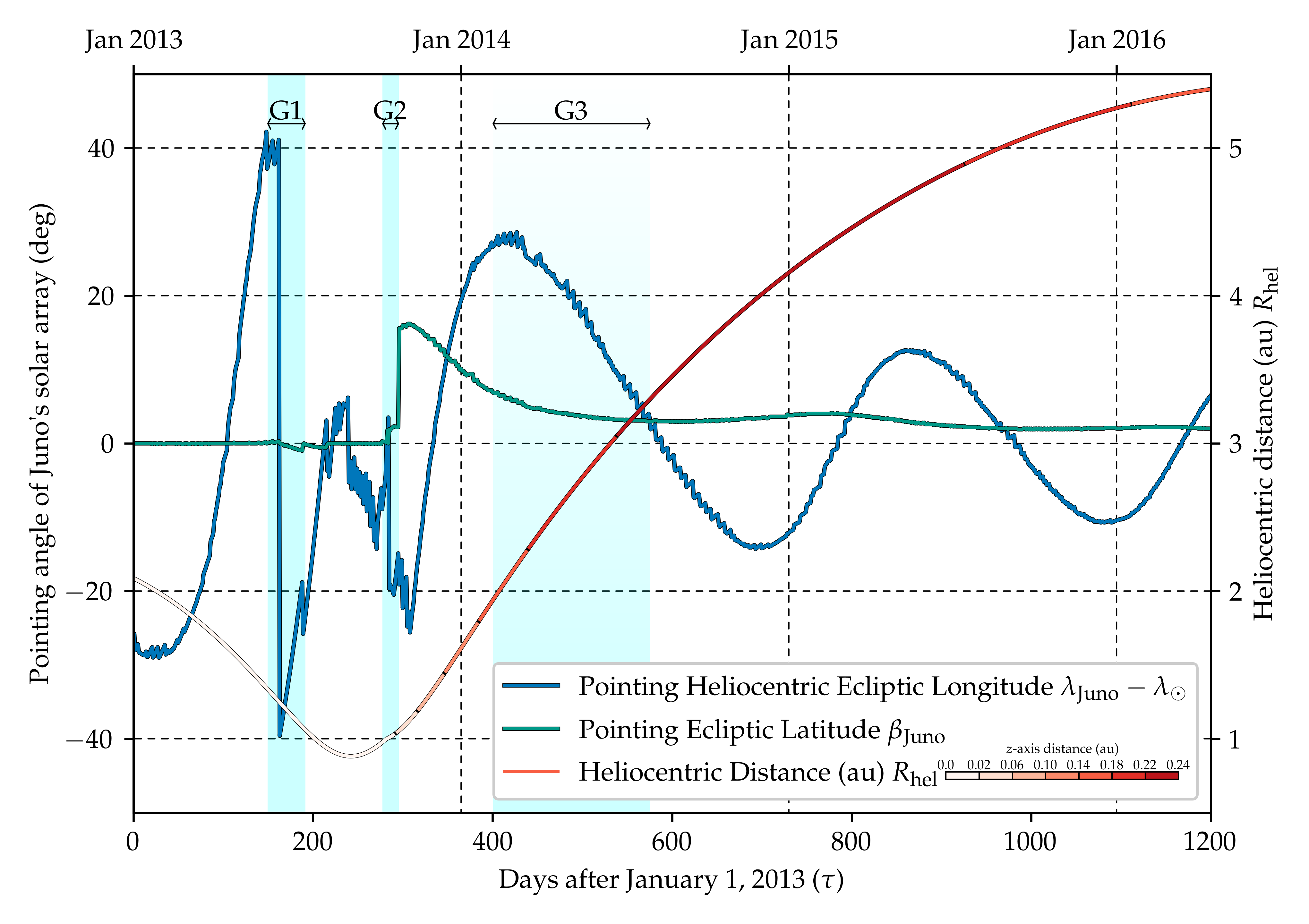}
\caption{Variations of the heliocentric distance $R_\mathrm{hel}$ of Juno and the pointing angles of the solar array $({\lambda_\mathrm{Juno}-\lambda_\odot}, \beta_\mathrm{Juno})$ in time starting on January 1, 2013. The primary $x-$axis shows the number of days since January 1, 2013 ($\tau$), whereas the secondary (top) $x$-axis shows the calendar years. The solar array pointing longitude ${\lambda_\mathrm{Juno}-\lambda_\odot}$ and latitude $\beta_\mathrm{Juno}$ in heliocentric ecliptic coordinates are represented by blue and green solid lines, respectively. The pointing angle values in degrees are denoted by the left-hand side $y$-axis. The spacecraft heliocentric distance $R_\mathrm{hel}$ in au is represented by the red/white solid line where its values are denoted by the right-hand side $y$-axis. The $R_\mathrm{hel}$ line is color coded by the distance of Juno from the ecliptic (in au), where the range of values is shown in the legend. The three cyan shaded areas represent time periods with limited/no detection rates. G1 and G2 data gaps are caused by a significant orientation change of the spacecraft. G3 represents a period when the instruments were in a different operational mode, during which impact detection was suppressed.}
\label{FIG:Juno_Geometry}
\end{figure}

In this article we analyze the data set of STEs recorded between January 1, 2013 and April 14, 2016 \citep{Jorgensen_etal_2021}. We denote the number of days after January 1st, 2013 as $\tau$, i.e. January 1st, 2013 is $\tau=0$ and April 14th, 2016 is $\tau=1199$. During this time, Juno was on its journey to Jupiter and experienced a multitude of orbit adjustment maneuvers. In this Section, we focus on the heliocentric distance of the spacecraft $R_\mathrm{hel}$ and the solar array pointing longitude and latitude $({\lambda_\mathrm{Juno}-\lambda_\odot}, \beta_\mathrm{Juno})$. The variations of $R_\mathrm{hel}$, $({\lambda_\mathrm{Juno}-\lambda_\odot}, \beta_\mathrm{Juno})$ over 1200 days starting January 1st, 2013 are shown in Figure \ref{FIG:Juno_Geometry}. The heliocentric ecliptic longitude ${\lambda_\mathrm{Juno}-\lambda_\odot}$ of the solar array pointing spans from $-40^\circ$ to $+40^\circ$ with an abrupt change at $\tau=160$ (denoted as G1 in Fig. \ref{FIG:Juno_Geometry}). After the close encounter with Earth on October 9, 2013 (denoted as G2, $\tau=281$), the pointing longitude shows damped oscillations as Juno flies through the main asteroid belt toward Jupiter. These large oscillations in ${\lambda_\mathrm{Juno}-\lambda_\odot}$ mean that the solar arrays are sensitive to different impact directions during the time period of our analysis. The pointing ecliptic latitude $\beta_\mathrm{Juno}$ is aligned with the ecliptic until $\tau=281$ (G2). After that, $\beta_\mathrm{Juno}$ increased to $16.2^\circ$ and then gradually decreased to $1.4^\circ$ at the beginning of January 2016.

Juno's heliocentric distance $R_\mathrm{hel}$ changes significantly from the closest point at $R_\mathrm{hel} = 0.88$ au at $\tau=242$ to $R_\mathrm{hel} = 5.44$ au at $\tau=1199$. In Fig. \ref{FIG:Juno_Geometry}, we also show the distance of Juno from the ecliptic ($z$-axis distance) using different color coding (shades of red) of $R_\mathrm{hel}$. Until the close approach to Earth (G2), Juno stays in the ecliptic {\added{($|z|<0.00024$ au)}} and then its distance from the ecliptic increases up to $0.24$ au at $\tau=720$.

In the STE data set, two time periods with data gaps (G1 and G2) and one with lower data collecting efficiency (G3) were identified \citep{Jorgensen_etal_2021}. G1 and G2 are correlated with significant spacecraft maneuvers and abrupt changes in spacecraft pointing. We denote G1 as the time period between $\tau\in[149,191]$ and G2 as the time period between $\tau\in[277,295]$. G3 is the time period that Juno spends roughly in the main belt and is represented by a cyan gradient in the time period between $\tau\in[400, 575]$.

In summary, Juno's trajectory and solar array's pointing during the analyzed time period are quite complex and undergo significant variations. Therefore, it is important to correctly assess the impact probability of particles in the meteoroid complex with Juno informed by the values shown in Figure \ref{FIG:Juno_Geometry}. For this purpose, we employ the \citet{Kessler_1981} method discussed in Section \ref{SEC:Collisional_Probability}. We emphasize that the initial analyses of these STEs assumed that all particles which could generate STEs impacted Juno followed a  near-circular low-inclination Keplerian orbit and did not account for the collision probability or detectability based on the impact directions of such particles to impact Juno \citep{Jorgensen_etal_2021}. These factors must be accounted for to attempt to use the STE dataset to infer the properties of the Zodiacal impact environment. In the following Section we show our analysis of Juno's expected dust impact profile accounting for these factors.

\section{Results - Meteoroid Model impacts on Juno solar arrays}
\label{SEC:MODEL_IMPACTS}
\begin{figure}
\epsscale{0.85}
\plotone{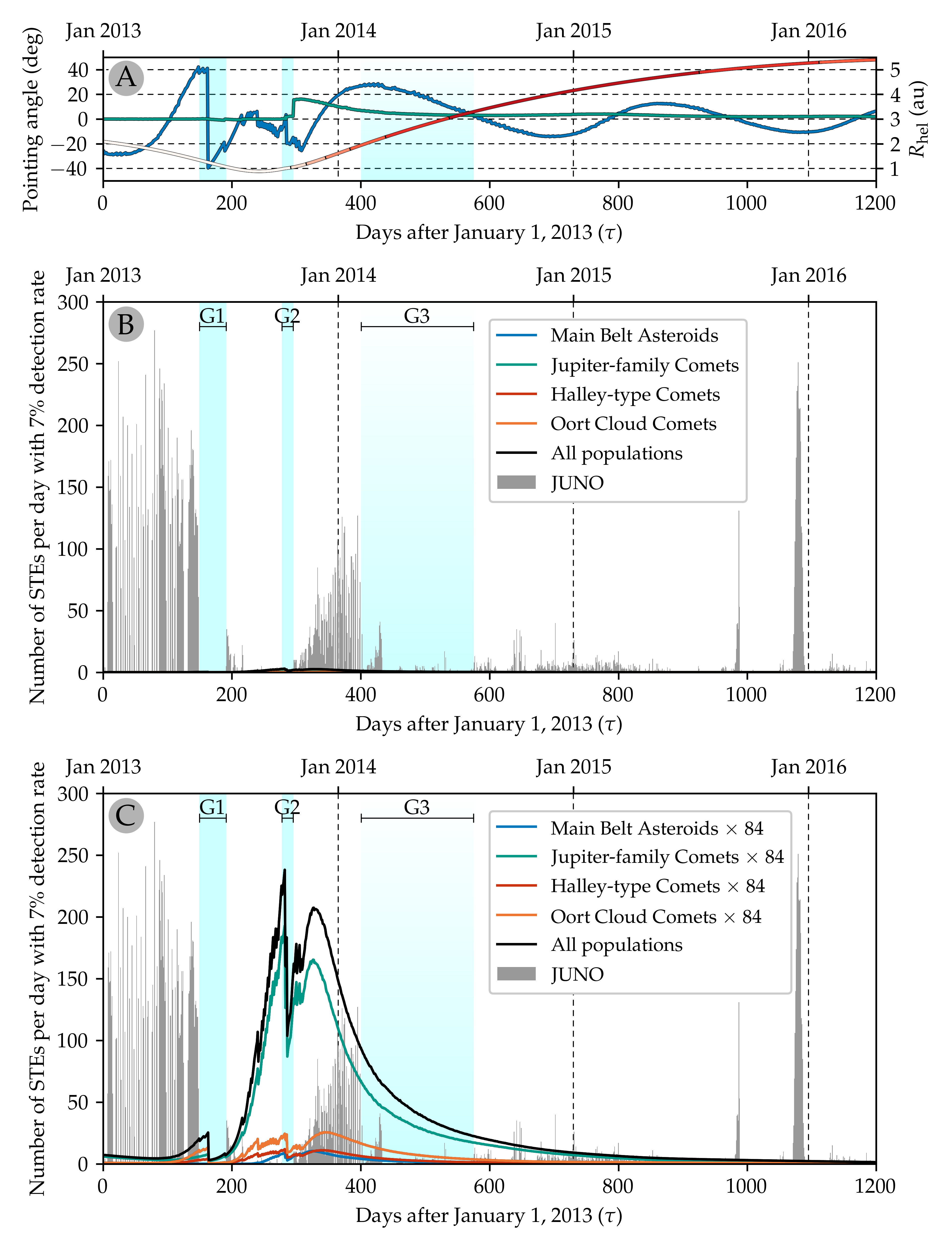}
\caption{\textit{Panel A:} The same as Fig. \ref{FIG:Juno_Geometry}. \textit{Panel B:} Number of star tracker events (STEs) on Juno's solar array per day between January 1, 2013 and April 14, 2016 (gray histogram). The number of STEs estimated for the four model meteoroid populations and their sum is shown as solid lines: main belt asteroids (blue solid line), Jupiter-family comets (green solid line), Halley-type comets (red solid line), Oort Cloud comets (orange solid lines), and their combined value (black solid line). The number of predicted meteoroid impacts is multiplied by the $\epsilon = 7\%$ detection rate estimated in \citet{Benn_etal_2017} and used in \citet{Jorgensen_etal_2021} to convert model impact rates to expected STEs. Note, that our meteoroid model predicts maximum $N=2.98$ detected impacts per day, which makes the solid lines very close to zero. We want to emphasize the magnitude of disagreement between the model and the observation. The three cyan shaded areas represent the time periods with limited/no detection rates. G1 and G2 data gaps due to significant orientation change of the spacecraft. G3 represents a period when the instruments were in a different operational mode during which STE detection was suppressed. This plot shows that our meteoroid models predict 2-3 orders of magnitude fewer STEs than the reported numbers.
\textit{Panel C:} The same as \textit{Panel B} but now all model expected STE rates are multiplied by a factor of 84 to scale the maximum predicted number of STEs to 250. Our model cannot reproduce the first 200 days of STE rates and shows different rate profile for the rest of the analyzed time period. See video version of this Figure that is also showing the impact direction variations for all meteoroid population in time in the online version (\url{https://iopscience.iop.org/article/10.3847/PSJ/ac4019})
}
\label{FIG:JUNO_Model_Impacts_Revision}
\end{figure}

First, we analyze the number of impacts per day our model predicts for the Juno spacecraft, assuming the detection cross-section of 60 m$^{2}$ pointing toward the angles shown in Fig. \ref{FIG:Juno_Geometry} and that each individual model impact generates a single STE with 7\% efficiency (Eq. \ref{EQ:IMPACT_RATE}). During the entire time frame of our analysis, we expect a peak rate of $N=2.98$ STEs per day for the sum of all four meteoroid populations investigated here (Fig. \ref{FIG:JUNO_Model_Impacts_Revision}B). The expected peak STE rate occurs during the data gap G2 period at $\tau = 280$ days, and a second maximum of $N=2.59$ impacts per day at $\tau = 325$ days. The total number of detected STEs during the 1200 day period is $N_\mathrm{tot} = 511.9$. From Figure \ref{FIG:JUNO_Model_Impacts_Revision}B, and the maximum number of STEs, we find that our meteoroid model cannot qualitatively or quantitatively reproduce the Juno STE rates \citep{Jorgensen_etal_2021} and provides STEs rates 1-2 orders of magnitude smaller. Note, that even assuming 100\% detection efficiency (scaling our rates by $\sim$14), our model predictions would still fall short both in terms of daily impact rates as well as the total number of STEs.


The meteoroid model used here has been successful to reproduce the meteoroid number flux for the Lisa Pathfinder spacecraft \citep{Thorpe_etal_2019}, which was sensitive to impacts with momentum $>1.0~\mu$Ns with 100\% efficiency. This translates to impactor diameters $D\approx 35-45~\mu$m for impact velocities between $V_\mathrm{imp}=10-20$ km s$^{-1}$ and particle bulk density $\rho = 2,000$ kg m$^{-3}$. The number of particles impacting the spacecraft is modulated by the size-frequency distribution, which is one of the free parameters of the meteoroid model. To test this free parameter, we assume that Juno is in fact detecting particles much smaller than $D=10~\mu$m or that the size-frequency distributions of meteoroid populations are steeper and produce more smaller meteoroids than the model predicts. To simulate this, we multiplied all model fluxes by a factor of{ \replaced{80}{84} to \added{increase the maximum predicted number of STEs to $N=250$, and to}} better illustrate the variability of the impactor flux on Juno's solar arrays with time over the entire analysis period (Figure \ref{FIG:JUNO_Model_Impacts_Revision}C). It is evident that all four meteoroid populations are expected to have peak STE rates at Juno during the G2 data gap and between $\tau=310$ and $\tau=330$. Moreover, the model STE rate time variation cannot reproduce the first 180 days after January 1, 2013 where $\sim100-250$ STEs per day on Juno's solar arrays were reported. Even if we re-scale any of the meteoroid populations to match the number of STEs before the data gap G1, the rest of the analyzed time frame would be inconsistent with the values reported for $\tau>200$ days, reaching $N=5,000$ predicted STEs per day between data gaps G2 and G3.

Figure \ref{FIG:JUNO_Model_Impacts_Revision}C shows how sensitively dependent the modeled STE rate is on the pointing of the solar arrays, as seen from the modulations of $N$ during spacecraft maneuvers during data gaps G1 and G2. Another factor that significantly modulates the impactor flux is the heliocentric distance of the spacecraft, since the meteoroid environment number density is proportional to heliocentric distance as $\propto R_\mathrm{hel}^{-1.3}$ \citep{Leinert_etal_1981,Stenborg_etal_2021}, and the orbital velocity of meteoroids scales as $\propto R_\mathrm{hel}^{-0.5}$ (see e.g., vis-viva equation). Approximately 150 STEs per day are recorded around $\tau = 10$ days, where Juno is at $R_\mathrm{hel} = 2$ au and no significant increase in STEs is observed during the inbound phase of spacecraft orbit. In fact, there are almost no STEs between the G1 and G2 data gaps, where Juno's $R_\mathrm{hel}$ reaches its minimum below 1 au. From a multitude of in-situ spacecraft data, we know there should be a considerable meteoroid flux of bound grains at and below 1 au \citep[e.g.][]{Grun_etal_1980,Szalay_etal_2021}, and we would not expect the Zodiacal Cloud to exhibit increasing density with increasing heliocentric distance. 

What causes the discord between the observation and the model? The most important factor is due to the anti-helion pointing of Juno's ``sensor" and the fact that during Juno's pre-perihelion passage ($\tau < 200$), meteoroids impact the spacecraft from the helion/sunward direction. This very same effect is observed and modeled for the Parker Solar Probe \citep{Szalay_etal_2020}, where both Parker Solar Probe and Juno are on highly eccentric orbits in the Zodiacal Cloud. During Juno's post-perihelion leg the impactor direction shifts to the anti-helion direction, which results in elevated impact rates that are accentuated by the heliocentric distance of the spacecraft. We discuss dynamical reasons for the model-observation disagreement in more detail in the Discussion section.

From our analysis, we conclude that our model cannot reproduce neither the number of Juno STEs per day nor the general shape of the number of STE variations with time.  In the next section we explore the alternative hypothesis of dust generated inside the Martian Hill sphere by either Mars or its moons \citep{Jorgensen_etal_2021} and predict the STE rates for Juno from this population. These results also apply to a more general population of dust producing sources sharing the orbital space with Mars.


\section{Impacts of dust generated by Mars and its moons}
\label{SEC:MARS_IMPACTS}
In Section \ref{SEC:MARS_MODEL} we discussed how we created the dust cloud generated either by Mars or its moons, i.e. dust leaving the Martian Hill sphere. It is important to note, that there is no evidence for a large abundance of dust generated by Mars or its moons based on multiple spacecraft observations. For example the MAVEN spacecraft observed dust particle impacts using impact plasma generated voltage spikes, but these dust impacts were coming from interplanetary space and were in accordance with nominal Zodiacal Cloud dust models \citep{Andersson_etal_2015}. Additionally, there is no evidence for dust activity at either of Mars' moons Phobos and Deimos \citep{Pabari_Bhalodi_2017}. We pursue the hypothetical Martian dust for the sake of completeness and to potentially find a missing piece of the dust complex which was proposed to explain the STE observations \citep{Jorgensen_etal_2021}.

We track the orbital evolution of particles of vastly different sizes, where four of our sizes $D=1.5,2.0,2.5,5.0~\mu$m are not detectable by Juno using the $D\ge10~\mu$m cut-off, and the remaining three sizes $D=10,~20,~50~\mu$m should be detectable by Juno observing methods \citep{Jorgensen_etal_2021}. Particles with $D\le2.5~\mu$m ($\beta \ge 0.383$) can be ejected from the Martian Hill Sphere to Jupiter-crossing orbits due to radiation pressure, which leads to a more complex dynamical evolution due to frequent interactions with Jupiter and its mean-motion resonances. Particles with $D\ge5~\mu$m experience a simple Poynting-Robertson drag induced decay in semimajor axis $a$ and eccentricity circularization ($e\rightarrow0$) and occasional trapping in one of many mean-motion resonances with terrestrial planets similar to dust particles released from the main belt \citep[see e.g.,][]{Sommer_etal_2020}. Mean-motion resonances temporarily trap migrating dust particles, which stall the particles from spiraling toward the sun and pump the particles' eccentricities, but ultimately do not play a major role in the global shape of the dust cloud generated from the Martian Hill Sphere. 

\begin{figure}
\plotone{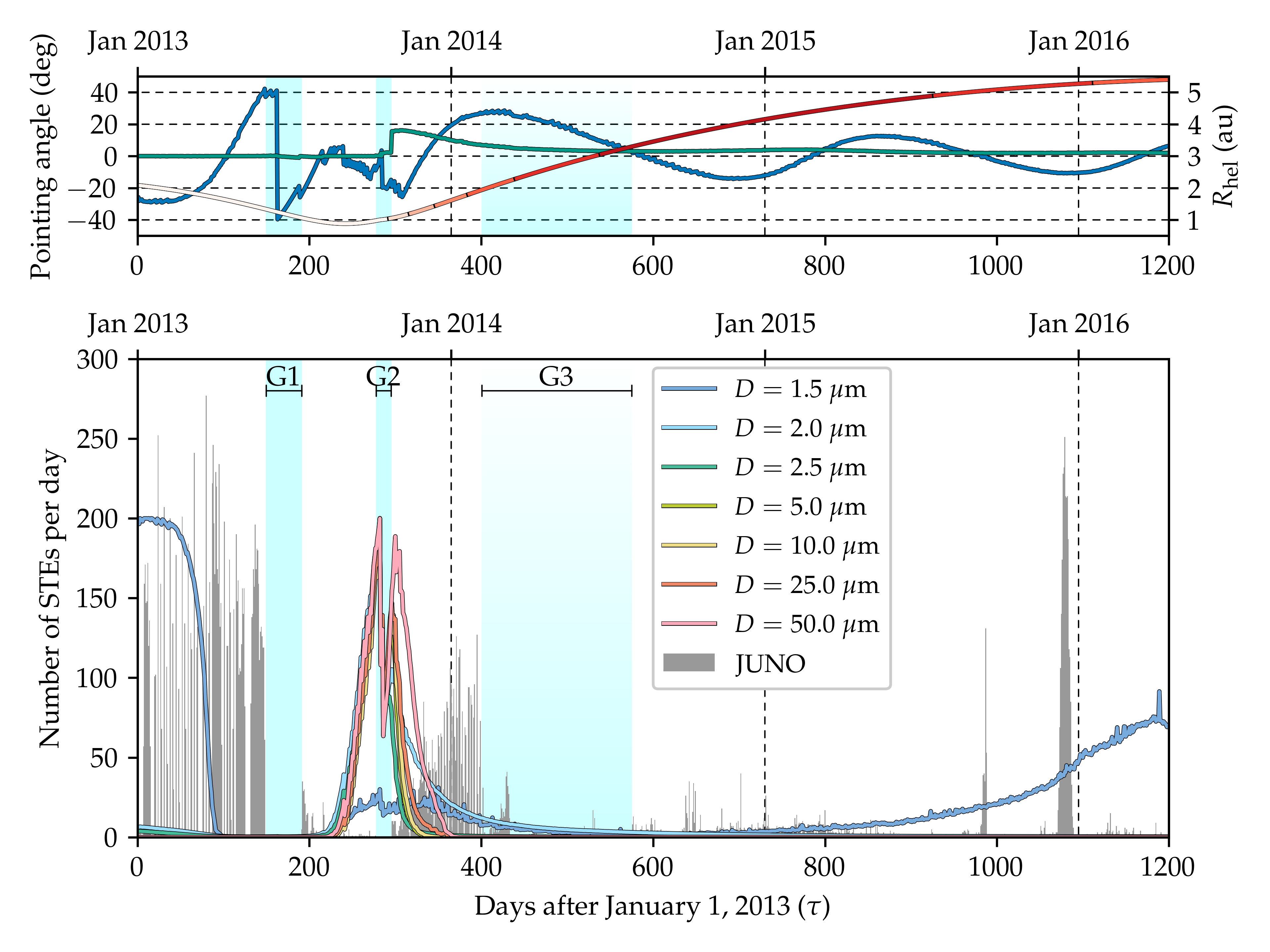}
\caption{The same as Fig. \ref{FIG:JUNO_Model_Impacts_Revision}, but now for dust particles released from the Martian Hill sphere. For each particle diameter, we scaled the according simulation to provide a maximum of $N=200$ \replaced{impacts}{modelled STEs} per day. Particle cloud with $D=1.5~\mu$m peak around January 1st, 2013, while the particle clouds \added{of larger particles} with $D>1.5~\mu$m have peak values \deleted{in mid-December 2013} \added{around the G2 gap} ($\tau\approx278$). \deleted{ whereas the maximum value of $N$ for $D=50.0~\mu$m occurs during the G2 data gap ($\tau = 278$).} No dust size or their combination is able to reproduce the STE rate profile reported in \citet{Jorgensen_etal_2021}. }
\label{FIG:JUNO_Mars_Dust}
\end{figure}

Figure \ref{FIG:JUNO_Mars_Dust} shows the variation in number of detectable impacts per day for 1200 days of the Juno mission starting January 1, 2013 for the model dust clouds generated from the Martian Hill Sphere. We scaled models of all particle sizes to provide the maximum number of STEs per day on Juno $N = 200$ for easier comparison of different particle sizes and STE rates. There are two categories of impact profiles for our Martian dust model. (A) Particles with $D = 1.5~\mu$m have peak values of $N$ at the beginning of our analysis time frame in January 2013 ($\tau = 0$) and decrease with time to $N=0$ around $\tau \approx 90$. The \deleted{three} smallest {\added{modelled} particle\added{s}} \deleted{diameters investigated here $D=1.5,2.0,2.5~\mu$m} start to impact Juno{ \added{again when Juno is close to its perihelion around the G2 gap and then again }}after the data gap G3, which corresponds to dust grains pushed by radiation pressure into orbits beyond the main belt to Jupiter-crossing orbits. (B) Meteoroids with $D\ge2.0~\mu$m do not get a large radiation pressure kick upon their ejection and their initial orbits are very similar to that of Mars. Most particles of these sizes do not impact Juno's solar array during the first 200 days, leaving the STE rates observed in this period {\replaced{mystery}{unexplained by any known or hypothetical dust population}}. \deleted{The largest meteoroids in our sample $D=50~\mu$m exhibit two peaks in $N$: one during the G2 data gap and a second one around mid-December 2013 ($\tau\approx345$). The second peak of $D=50~\mu$m particles is also the only significant impact period of $D=10,25~\mu$m particles from the Martian Hill Sphere. The rest of the STE rates before the G3 data gap reported for Juno is not reproduced by our model for Martian dust. }
{\added{All meteoroids with $D\ge2.0~\mu$m exhibit very similar impact rate profiles with peak values occurring during or just before/after the G2 data gap ($\tau \approx 278$) reflecting the expected particle number density increase closer to the Sun. The sudden shifts in modelled $N$ during the G2 data gap reflect the abrupt movements of Juno's solar array and the change of the detector pointing. Comparison to Fig. \ref{FIG:JUNO_Model_Impacts_Revision}C shows that the dust released from Martian Hill Sphere follows similar dynamical pathways to meteoroids originating from Jupiter-family comets and main belt asteroids.}}

We furthermore analyzed a broader range of diameters for dust ejected from the Martian Hill Sphere, which is not shown in Figure \ref{FIG:JUNO_Mars_Dust}. We could not find any particular particle size or their combination that would even remotely reproduce the Juno STE rate profile. For this reason, our analysis cannot support the Martian dust hypothesis due to the basic disagreement of our model with the STE observations.{\added{We thus conclude that the Martian dust hypothesis presented by \citet{Jorgensen_etal_2021} is invalid as an explanation to the STE observations.}}

\section{Discussion}
In the previous Sections we showed that no combination of our meteoroid models for the most abundant inner solar system populations can reproduce the Juno star tracker event (STE) rate profile \citep{Jorgensen_etal_2021}. We also showed that dust ejected from the Martian Hill Sphere of various sizes cannot reproduce the Juno STE profile either. This brings us to the conclusion that either (A) our meteoroid models or our methods are vastly incorrect or (B) that STE observations do not represent the record of {\added{individual}} dust impacts but rather a detection of a{ \added{dust impact}} phenomenon that is {\added{either more complex in nature or not related to dust impacts at all.}} \deleted{much more complex in nature.} 
Setting aside the fact that meteoroid models used in this article are able to reproduce most of the meteoroid related phenomena in the inner solar system with high fidelity, \deleted{the rules of orbital dynamics play against} the hypothesis that Juno STE rates are direct detections of impacts from the Zodiacal Cloud \citep{Jorgensen_etal_2021}{ \added{are in direct conflict with the rules of orbital dynamics}}. During the first 250 days after January 1, 2013 Juno was plunging to the inner solar system on a highly eccentric orbit ($e=0.44$). During this pre-perihelion phase, Juno's orbital vector was closer to the Sun than the orbital velocity vector of an object on a circular orbit and thus the sunward side of the spacecraft experienced enhanced meteoroid fluxes. However, Juno STEs were exclusively tied to the anti-helion side of the spacecraft, where impacts were greatly diminished. On the other hand, when Juno was in the post-perihelion phase, the number of impacts from the anti-helion direction was enhanced and we see the enhancement of expected impact rates in Figure \ref{FIG:JUNO_Model_Impacts_Revision}. The same effect is expressed in our Martian dust model on bound orbits as shown in Fig. \ref{FIG:JUNO_Mars_Dust}. For this purely dynamical reason, any bound population of meteoroids orbiting the Sun will have higher flux on the anti-sunward side of the solar panels during Juno's post-perihelion passage phase. This is, however,{ \replaced{incompatible}{in disagreement}} with the Juno STE rates \citep{Jorgensen_etal_2021}. 

\begin{figure}
\epsscale{0.95}
\plotone{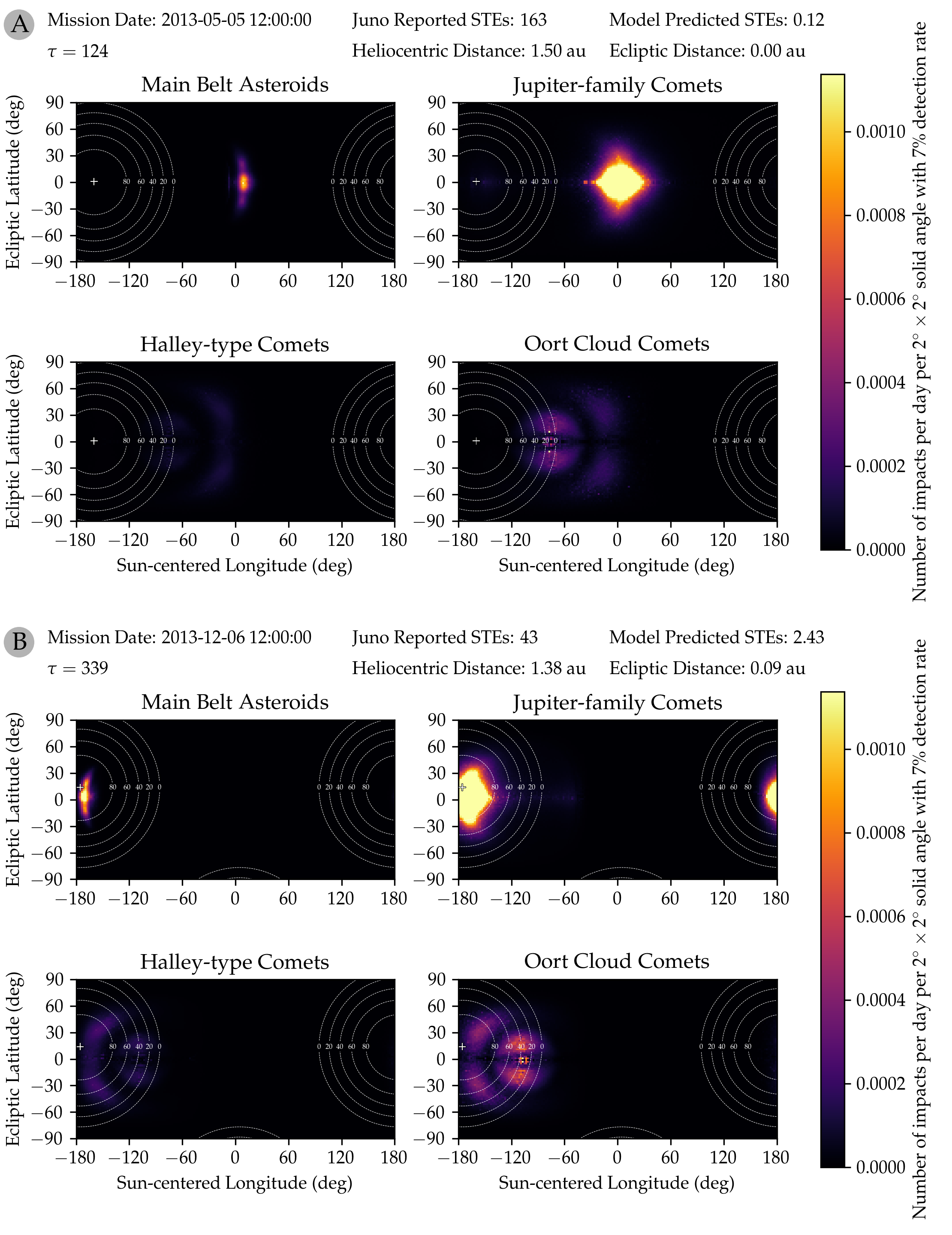}
\caption{Meteoroid impact directions of the four model populations. \textit{Panel A}: Number of impacts per day per $2^\circ \times 2^\circ$ solid angle in sun-centered ecliptic coordinates for MBA, JFC, HTC, and OCC model meteoroids (color coded). We assume that the target is a sphere with a cross-section of 60 m$^2$ and the impact detection efficiency 7\%. Juno's solar array pointing direction is represented by a cross symbol that is surrounded by dashed circles showing levels of $\cos\varphi$ factor (Eq. \ref{EQ:Collecting_Area}); i.e. the attenuation of the effective collecting area due to shallower impact angles. Any impacts outside the "0" contour result in no STEs. This panel shows impactor distribution on May 5, 2013 ($\tau=124$) when Juno reported 163 STEs and our model predicted $N=0.12$ STEs. The spacecraft was at heliocentric distance $R_\mathrm{hel}=1.5$ au, moving toward perihelion. Most of the impactors striked Juno from sunward direction and are not expected to produce STEs. 
\textit{Panel B}: The same as Panel A but now for the post-perihelion passage on December 6, 2013 ($\tau=339$). The spacecraft was at $R_\mathrm{hel}=1.38$ au, Juno reported 43 STEs, our model predicted $N=2.43$ STEs. All meteoroid populations were concentrated in the anti-helion region and were impacting Juno's solar array efficiently. } \label{FIG:JUNO_Impact_Directions}
\end{figure}

We show this pre- and post-perihelion impact direction shift for all four meteoroid populations in Fig. \ref{FIG:JUNO_Impact_Directions}, where we focus on two temporal snapshots ($\tau=125$ and $\tau=340$). In Fig. \ref{FIG:JUNO_Impact_Directions}A Juno is in the pre-perihelion phase and most of the impacts on the spacecraft are concentrated in the sunward direction (ecliptic longitude $\lambda-\lambda_\odot = 0^\circ$). The concentric rings representing different levels of $\cos \varphi$ show that only OCC meteoroids are able to impact the anti-sunward facing side of the solar array, though on very shallow angles, decreasing the collecting area significantly (Eq. \ref{EQ:Collecting_Area}). In the post-perihelion phase (Fig. \ref{FIG:JUNO_Impact_Directions}B) impacts from all populations shift to the anti-sunward direction and are able to generate STEs; however, still with much smaller rates than the number of STEs reported for Juno. {\added{We additionally tested hypothetical detector pointings such as the helion/sunward pointing detector, detector pointing into the ram direction of the spacecraft (the direction of Juno's instantaneous velocity vector), or the anti-ram direction. None of these hypothetical detector pointings were able to at reproduce the reported number of STEs or their variations in time.} }
We also include a movie (Fig. \ref{FIG:JUNO_Model_Impacts_Revision}) showing the distributions of impact directions for all four meteoroid populations, together with the STE rates reported in \citet{Jorgensen_etal_2021} and Juno's trajectory information for all 1200 days of our analysis. 

This leads us to the second scenario (B), suggesting that STEs {\replaced{might not be}{do not}} represent { \added{individual}} meteoroid impact events, but are rather records of \deleted{a} more complex {\added{impact}} phenomena{ \added{or events unrelated to meteoroid impacts}}. An extensive exposure of International Space Station (ISS) solar array to meteoroid impacts showed the variety of effects these impacts can have on the structure and function of space-borne solar arrays \citep{Hyde_etal_2019}. 
Without proper laboratory experiments, we cannot rule out the possibility that singular impacts can cause various cascade effects which can result in subsequent material ejection events from the solar array. Such material ejection could occur as a result of continuous erosion by much smaller dust grains than those considered to cause the reported impacts, or thermal stress \citep{Wienhold_Persons_2003}. These events could be delayed with respect to the original impact and would not be related in the STE dataset to the original impact. This would also explain the existence of gaps in the first 180 days after January 1, 2013, where on certain days \citet{Jorgensen_etal_2021} reports 100+ STEs per day (i.e., total of 1000+ impacts per day considering 7\% STE detection efficiency), while on subsequent days there are no impacts recorded at all. This kind of behavior does not follow the expected Poisson statistics seen for other meteoroid related phenomena, such as impacts on other spacecraft \citep[e.g.,][]{Page_etal_2020, Szalay_etal_2021,  Pusack_etal_2021}, meteor detections at Earth \citep[e.g.,][]{Pokorny_Brown_2016,Jenniskens_etal_2020}, or responses of airless bodies to meteoroid impacts \citep[e.g.,][]{Burger_atal_2014,Szalay_etal_2015}.{ \added{For an overview of the observations and modelling of meteoroid related phenomena, see the recent review work by \citet{Janches_etal_2021}}.}

\section{Conclusions}
In this article, we showed that currently existing models for the meteoroid environment in the inner solar system cannot reproduce the star tracker events observed during Juno's interplanetary cruise phase \citep{Jorgensen_etal_2021}. Both the number of expected STEs Juno's solar array should experience (Fig. \ref{FIG:JUNO_Model_Impacts_Revision}B) and the meteoroid impact variations with time (Fig. \ref{FIG:JUNO_Model_Impacts_Revision}C) do not show any potential to reproduce the reported STEs. 

We also showed that a hypothetical population of dust and meteoroid particles ejected from the Martian Hill Sphere is not capable of reproducing the observed STE rates reported \citep{Jorgensen_etal_2021}. Neither bound nor unbound dust grains from this hypothetical Martian population show any dust impact profile signatures that could explain the STEs if these would indeed be generated by  dust impacts(Fig. \ref{FIG:JUNO_Mars_Dust}).

Ultimately, we showed that the orbital dynamics prefer impacts from the sunward direction in the first 200 days after January 1, 2013, while the observed STE rate is 100+ per day from the anti-sunward direction. This would imply much higher fluxes in the post-perihelion passage phase of Juno's voyage, which is not reflected in the STE data. 

Unless there exists an unknown and dense population of meteoroids that shows increasing spatial density with increasing heliocentric distance, contrary to all other Zodiacal Cloud observations to date, and is specifically tuned to impact Juno during the first 200 days after January 1, 2013, we have to conclude that the Juno STE events are in fact not records of individual impacts of meteoroids, but rather records of a complex phenomenon observable by Juno's star trackers that may or may not be related to meteoroid impacts. 

\software{
\texttt{gnuplot} (\url{http://www.gnuplot.info}) $\bullet$
\texttt{swift} \citep{Levison_Duncan_2013} $\bullet$ \texttt{matplotlib} (\url{https://www.matplotlib.org}) \citep{Hunter_2007} $\bullet$
\texttt{SciPy} (\url{https://www.scipy.org}) $\bullet$ \texttt{SPICE} \citep{Acton_etal_2018}
}

\facilities{NASA Center for Climate Simulation (NCCS), NASA Advanced Data Analytics PlaTform (ADAPT)}

\begin{acknowledgments}
\noindent Funding: P.P. and M.J.K. were supported by NASA ISFM EIMM award, the NASA Cooperative Agreement 80GSFC21M0002, as supported by NASA Solar System Workings award No. 80NSSC21K0153 . P.P., J.R.S, and M.H. were supported and NASA SSERVI award 80NSSC19M0217 \\
Author contributions: \\
P.P.: Conceptualization, formal analysis, funding acquisition, investigation, methodology, software, validation, visualization, writing - original draft, writing - review \& editing \\
J.R.S: Conceptualization, formal analysis, investigation, visualization, writing - original draft, writing - review \& editing\\
M.H.: Conceptualization, formal analysis, funding acquisition, investigation, writing - original draft, writing - review \& editing \\
M.J.K.: Conceptualization,  writing - original draft, writing - review \& editing\\
Competing interests: Authors declare no competing interests.\\
Data and materials availability: All data is available in the main text or the supplementary materials.
\end{acknowledgments}

\end{document}